\documentclass[a4paper,superscriptaddress,twocolumn,aps,prb,showpacs,longbibliography]{revtex4-2}
\usepackage{dcolumn}
\usepackage{bm}
\usepackage[colorlinks=true,citecolor=blue,linkcolor=blue,pdfhighlight =/O]{hyperref}
\usepackage{graphicx}
\usepackage{amsmath}
\usepackage{cancel}
\usepackage{dsfont}
\usepackage{mathtools}
\usepackage[dvipsnames]{xcolor}
\usepackage{soul}
\usepackage[normalem]{ulem}
\usepackage{bm}
\usepackage{amsmath,amssymb,amsfonts,latexsym,fancyhdr,graphicx}
\usepackage{simplewick}
\usepackage[english]{babel}
\usepackage{graphicx}
\usepackage{subfigure}
\usepackage{xcolor}
\usepackage{color}
\usepackage{verbatim}
\usepackage{units}
\usepackage{pdfpages}

\makeatletter
\AtBeginDocument{\let\LS@rot\@undefined}
\makeatother

\hypersetup{urlcolor=blue}%

\begin{document}
\title{Simulating bistable current-induced switching of metallic atomic contacts by electron-vibration scattering}
\author{Markus Ring}
\affiliation{Institute of Physics, University of Augsburg, 86159 Augsburg, Germany}
\affiliation{Department of Physics, University of Konstanz, 78457 Konstanz, Germany}
\author{Fabian Pauly}
\affiliation{Institute of Physics, University of Augsburg, 86159 Augsburg, Germany}
\author{Peter Nielaba}
\affiliation{Department of Physics, University of Konstanz, 78457 Konstanz, Germany}
\author{Elke Scheer}
\affiliation{Department of Physics, University of Konstanz, 78457 Konstanz, Germany}
\email[]{elke.scheer@uni-konstanz.de}

\date{\today}

\begin{abstract}
We present a microscopic model, describing current-driven switching in metallic atomic-size contacts. Applying a high current through an atomic-size contact, creates a strong electronic nonequilibrium that excites vibrational modes by virtue of the electron-vibration coupling. Using density functional theory (DFT) in combination with the Landauer-Büttiker theory for phase-coherent transport, expressed in terms of nonequilibrium Green's functions (NEGFs), we study the current-induced forces arising from this nonequilibrium and determine those vibrational modes which couple most strongly to the electronic system. For single-atom lead (Pb) contacts we show specific candidates for bistable switches, consisting of two similar atomic configurations with differing electric conductance. We identify vibrational modes that induce a transition between these configurations. Our results reveal a possible origin of bistable switching in atomic-size contacts through excitation of vibrations by inelastic electron scattering and underline the power of the combined DFT-NEGF approach and statistical mechanics analysis of a Langevin equation to overcome the time-scale gap between atomic motion and rare switching events, allowing for an efficient exploration of the contacts' configurational phase space.
\end{abstract}

\pacs{}

\maketitle

\section{Introduction}

Bistable atomic-scale conductance switches are considered as possible building blocks for nanoelectronic circuits \cite{Evers2020}. In a two-terminal configuration and activated by controlled electromigration they are ultimately miniaturized \cite{Schirm2013}. The term electromigration denotes the rearrangement of atoms inside a conductor in response to an applied bias voltage or flowing charge current. Electromigration in macroscopic conductors is reported to be a thermally driven process by the dissipated Joule heat \cite{Hoffmann2017}. While atomic-size switches are straightforward to realize experimentally, the microscopic theory is involved. Electromigration requires the description of the coupled electronic and atomic motion, which is typically separated along the lines of the Born-Oppenheimer approximation due to the large mass difference between electrons and atoms. The treatment of current-induced atomic rearrangements in a junction hence requires in principle complex dynamics simulations, bridging electronic and atomic time scales of several orders of magnitude. 

Metals can sustain high current densities, and electromigration is a relevant mechanism for atomic rearrangements \cite{Debenedetti2001}. Different models of electromigration on the atomic scale have been suggested , including the excitation of local vibrational modes due to inelastic scattering of electrons \cite{Lue2010, Todorov2014, Lue2015}. These inelastic scattering events cause forces that act on the atoms \cite{Lue_sgle_2019, Lue2020}. Although the microscopic processes are in principle clear, their implementation in molecular dynamics approaches proves difficult, since forces are nonconservative \cite{Dundas2009}. Recent theoretical work addressed this problem with ab-initio molecular dynamics, including heating through nonconservative forces and identifying hot spots and vibrational modes especially excited by the electronic nonequilibrium \cite{Lue2020}. 

The study of switching is challenging because of the large difference in time scales in the mechanics of interest. The typical time scale for atomic thermalization is picoseconds, while electronic relaxation happens much faster within femtoseconds. Even on the picosecond  time scale, however, major atomic relocations, causing electrical switching events, are rare. They typically happen in the microsecond range, as determined by the measurement resolution of experimental setups. This difference in time scales of some 9 orders of magnitude from the femtosecond, necessary to resolve the electronic subsystem, to the microsecond, relevant for switching events, is the central obstacle in simulating electromigration of metallic atomic contacts. 

In the present work we bridge the time-scale gap by first integrating the electronic dynamics into effective forces on the atomic scale and then investigating the long-term limit of atomic dynamics. Our current-induced-forces approach identifies those vibrational modes that couple strongest to the electronic nonequilibrium. Utilizing these vibrations to evolve the contact configuration can lead to different local minima in the configurational phase space. For a contact with $N$ flexible atoms, this strategy reduces the dimensionality of the search space of other stable configurations from 3$N$ to $O(1)$. Consequently, possible stable contact geometries are found in a computationally efficient way. 

We use the established formalisms of DFT and Landauer-Büttiker scattering theory to describe the phase-coherent electron transport, expressing the transport in terms of NEGFs. The inelastic scattering of electrons by vibrations of the system is taken into account in a time-averaged fashion through current-induced forces in a Langevin equation for the atoms, with a nonconservative friction kernel taking into account the nonequilibrium electron bath. The Langevin equation for the displacements $\vec{x}$ of all atoms from their equilibrium positions has the form
\begin{equation}
    \mathbf{m}\cdot\ddot{\vec{x}} + \boldsymbol{\eta}(V)\cdot \dot{\vec{x}} + \mathbf{D}(V)\cdot \vec{x} =  \vec{f}(V), \label{eq:Langevin}
\end{equation}
where $\mathbf{m}$ is the diagonal matrix of all atomic masses. The dynamical matrix $\mathbf{D}(V)$, the friction matrix $\boldsymbol{\eta}(V)$ and the random force $\vec{f}(V)$ are perturbed by the electronic nonequilibrium, as represented by the indicated dependence on the voltage $V$ \cite{Lue2010}. This perturbation adds antisymmetric contributions to the voltage-dependent matrices in Eq.~(\ref{eq:Langevin}), which lead to nonconservative forces. 

Analysis of the Langevin equation (\ref{eq:Langevin}) allows to determine threshold voltages, when specific excited vibrations become effectively undamped \cite{Lue2010}. The collective motion of the atoms along these modes is a potential mechanism for a switching process, since the undamped vibrations can lead to a mechanical instability of the contact configuration. In a previous work \cite{Ring2020} we computed threshold voltages for metallic atomic junctions of four different elements and compared them in a statistical analysis to experimentally extracted switching voltages. The good agreement between both corroborates vibrational pumping as possible switching mechanism. The present work is devoted to identifying bistable, i.e., reversible bivalued, switching processes as well as the underlying collective atomic motion based on this mechanism of electronic-vibrational excitations.

\begin{figure*}
    \centering
 \includegraphics[scale=0.8]{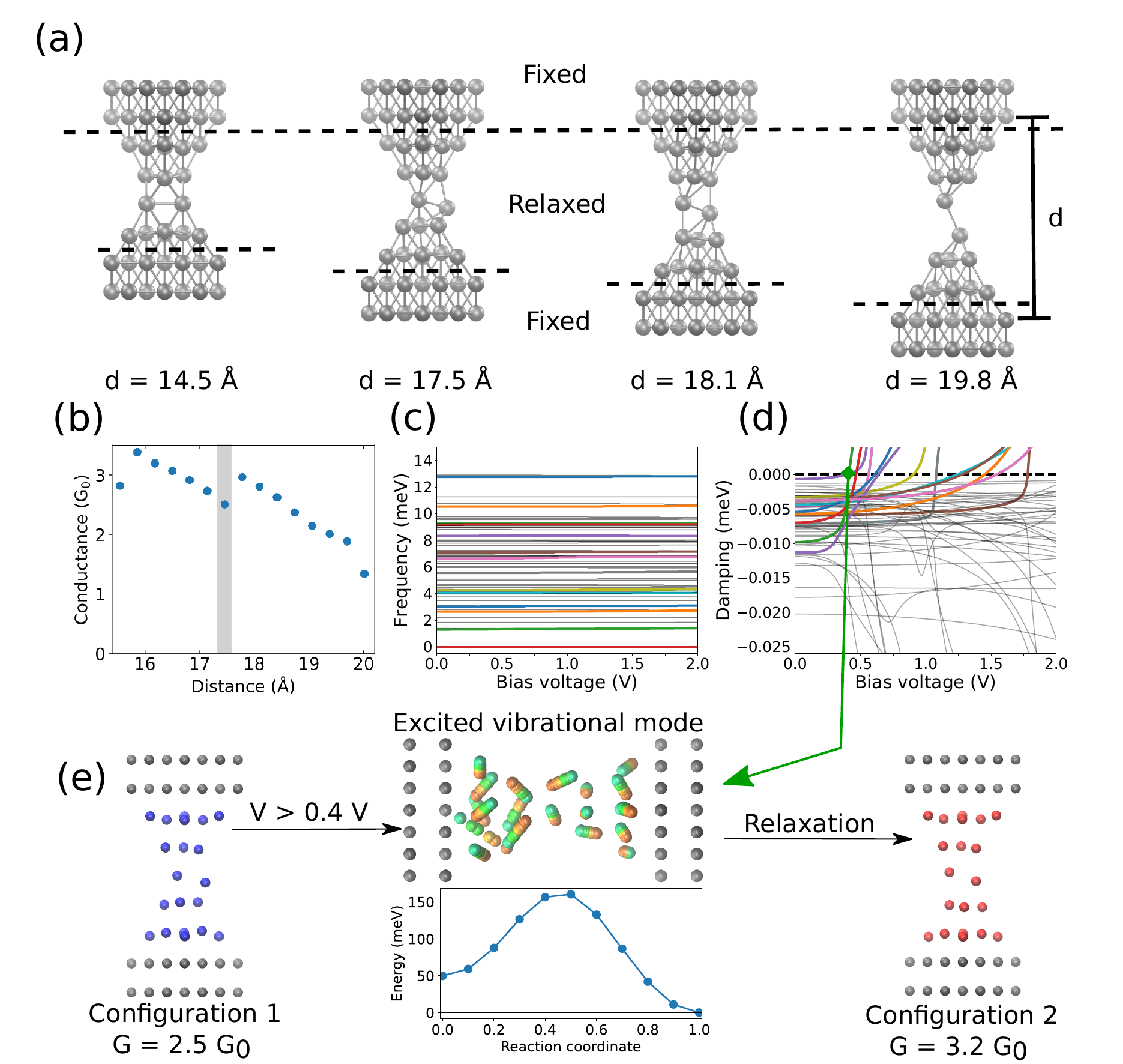}
    \caption{Scheme of the simulation process for describing current-induced atomic rearrangements based on electron-vibration coupling. (a) Contact geometries at various electrode separations $d$. Fixed and relaxed atoms are separated by dashed lines. (b) Conductance as a function of the electrode separation $d$. The contact at $d=17.5$~\AA, which is studied further in panels (c-e), is marked in gray. (c) Vibrational frequencies as a function of the applied bias voltage for the contact at $d=17.5$~\AA. (d) Damping of the vibrations as a function of the bias voltage $V$ for the contact at $d=17.5$~\AA.  (e) Starting configuration 1 of the atomic contact (left, blue) at $d=17.5$~\AA; displacement of its atoms by the mode with the lowest threshold voltage (middle), requiring $V>0.4$~V to become undamped; contact configuration 2 (right, red) after a corresponding relaxation, i.e. energy optimization of atomic positions. The resulting junction configurations 1 and 2 exhibit different conductance of $2.5G_0$ and $3.2G_0$, as indicated in the figure. The motion of atoms in the central relaxed junction part for one period of the unstable vibration is shown by snapshots in green and orange. The connection to the calculated bias-dependent damping in panel (d) is indicated by a green arrow. In the middle panel, the energy barrier between the two configurations is shown as a function of the reaction coordinate $r$.  \label{fig:flowchart}}
\end{figure*}

\section{Computational Procedures}
\label{sec:Methods}
In order to describe a bistable electrical switching process, it is necessary to first identify the stable geometries of the switch and then a mechanism to transition between them. The simplest description of the switching is given by a reaction coordinate connecting these two states over an energy barrier in between. Here we use simulations to determine all of these aspects: We identify two states, find a process to transition between them and determine energy barriers. The simulation approach is summarized in Fig.~\ref{fig:flowchart}. 

In this work, we study atomic-size metallic contacts of Pb. Extended central clusters \cite{Pauly2008}, containing the central narrowest constriction and part of the electrodes, consist of around 60 atoms, out of which 20 can move freely between two slabs of Pb atoms, fixed in a crystalline structure at a predefined distance, see Fig.~\ref{fig:flowchart}(a).  
The distance $d$ between the first fixed electrode layers on both sides of the contact is set to values between \unit[15-20]{\AA} in $15$ steps of about $\Delta=\unit[0.3]{\text{\AA}}$, see  Fig.~\ref{fig:flowchart}(a) and \ref{fig:flowchart}(b). The movable atoms between the two fixed crystalline layers are then relaxed to their energetic minimum. We calculate electronic and vibrational structures as well as the electron-vibration coupling with the quantum chemistry software package TURBOMOLE \cite{turbomole, Turbomole2020, Buerkle2013}. In the calculations presented here in the main text, we use the def-SV(P) basis set \cite{Schaefer1992,Eichkorn1997}. Results for the def-TZVP basis set \cite{Schaefer1994,Eichkorn1997} are discussed in the Supplemental Material.  The properties are then used in the NEGF framework to calculate the energy-dependent electronic transmission function $\tau(E)$ and all the matrices needed in the Langevin equation (\ref{eq:Langevin}) \cite{Lue2010}. Conductance values are determined in the phase-coherent elastic approximation in the low-temperature limit as $G=G_0\tau(E_\text{F})$, with the conductance quantum $G_0=2e^2/h$ and the Fermi energy $E_\text{F}$. In the charge transport calculations we use $32 \times 32$ transverse $k$-points.  We have extended a code to calculate inelastic electron tunneling spectra \cite{Buerkle2013, frederiksen_inelastic_2007, McEniry_2008} to include current-induced forces, following the approach of L{\"u} \emph{et al.} \cite{Lue2010}. We Fourier transformed Eq.~(\ref{eq:Langevin}) to compute vibrational eigenvalues and eigenmodes for different voltages at a specific $d$, see Fig.~\ref{fig:flowchart}(c). Above a certain voltage, some modes reveal a sign change of the damping from negative to positive, see Fig.~\ref{fig:flowchart}(d), indicating that they become undamped and are enhanced in amplitude instead of decaying over time. We term these vibrational modes "runaway modes" and the respective voltages "threshold voltages". We suggest that these undamped vibrations trigger atomic rearrangements, see Fig.~\ref{fig:flowchart}(e).

To realize a bistable atomic switch, we are interested in pairs of contact configurations, which give rise to different electronic conductance for the same distance $d$ between the electrodes. To find such pairs of geometries, we mechanically manipulate a contact by compressing or stretching, see Fig.~\ref{fig:flowchart}(a). The corresponding conductance-distance trace exhibits features which are known from experiment, like conductance plateaus and abrupt jumps in between at atomic rearrangements \cite{Agrait2003,Schirm2013,Ring2020,Weber2018}, see Fig.~\ref{fig:flowchart}(b). At the distances at which jumps in conductance occur, a hysteresis with respect to reversing the direction of the distance change can be expected, and hence two different metastable configurations for the same $d$. Since the conductance-distance trace in Fig.~\ref{fig:flowchart}(b) actually arises from a compression, we took the contact geometry at a subsequent reduced distance step $d-\Delta$, stretched it by $\Delta$ and optimized atomic positions again. Starting from the initial blue points, shown in Figs.~\ref{fig:flowchart}(b) and \ref{fig:3plots}(a), this mechanical cycle of $\mp\Delta$ generates the red points in Fig.~\ref{fig:3plots}(a). With this mechanical manipulation approach, we identify pairs of configurations, called configurations 1 and 2, see Fig.~\ref{fig:flowchart}(e), with different conductance for the same distance. Points of bistability, marked in Fig.~\ref{fig:3plots}(a) by gray bars, identify candidates for bistable atomic switches, which may be either operated mechanically by stretching and compressing or by current-induced forces \cite{Schirm2013}. 

Let us now discuss, if the transition between the configurations 1 and 2 at a certain $d$ can be mediated by a runaway mode and what the energy barrier for the transition is, see Fig.~\ref{fig:flowchart}(e). For candidate structures to act as reversible bivalued switches, several additional conditions must be met. At first the identified configurations 1 and 2 need to be separated by an energy barrier. A barrier is necessary to prevent random switching that would be observed in experiment either as telegraph oscillations of the conductance or as a weighted average conductance if the switching time is faster than the experimental measurement resolution. We use a linear interpolation of all atomic coordinates between the initial and final configurations 1 and 2 as the reaction path. The reaction coordinate $r$ is thus defined by $\vec{x}_r = \vec{x}_1 + r \cdot (\vec{x}_2 - \vec{x}_1)$ with $r\in[0,1]$, where $\vec{x}_1$ and $\vec{x}_2$ denote initial and final positions, respectively. DFT calculations for different $r$ then show the presence or absence of a reaction barrier and quantify its size. We note that the linear interpolation will yield an upper energy barrier for the transition between initial to final states. The calculations allow us to decide, if the structures are sufficiently stable with regard to switching over a certain time at a given temperature, see Fig.~\ref{fig:flowchart}(e). Finally, concerning the transition, we compute the runaway modes of configurations 1 and 2, see Fig.~\ref{fig:flowchart} (c) and \ref{fig:flowchart}(d). At or above the threshold voltage, atomic motion along these undamped modes requires vanishing energy cost. Accordingly, we vary the contacts by moving the atoms along these modes, assuming amplitudes of $\pm1$, $\pm2$, $\pm4$ or $\pm8$ times the normalized vibrational eigenvector. The contact geometries, obtained by this displacement procedure from configuration 1, are relaxed again to find a local energetic minimum. If this new configuration agrees with configuration 2, we have identified a current-induced mechanism for a vibrational transition between states 1 and 2, as illustrated in Fig.~\ref{fig:flowchart}(e). We attempt the same for the transition from 2 to 1. If runaway modes are found that establish transitions in both directions, a reversible bistable switch has been detected.

\section{Results and discussion}
From the extended list of requirements, namely, to find two contact geometries at a specific $d$ with largely different conductance, current-induced vibrational transitions and a sufficiently high energy barrier between them, it comes clear that only a fraction of the simulations returns current-driven bistable switches. In our present study we found three candidates out of $30$ contact structures, obtained by mechanical manipulation at different $d$. In the following we describe a mechanical compression curve that contains the three candidates, out of which only one resulted in a vibrationally-driven switch.  

\begin{figure}
\includegraphics[width=0.8\columnwidth]{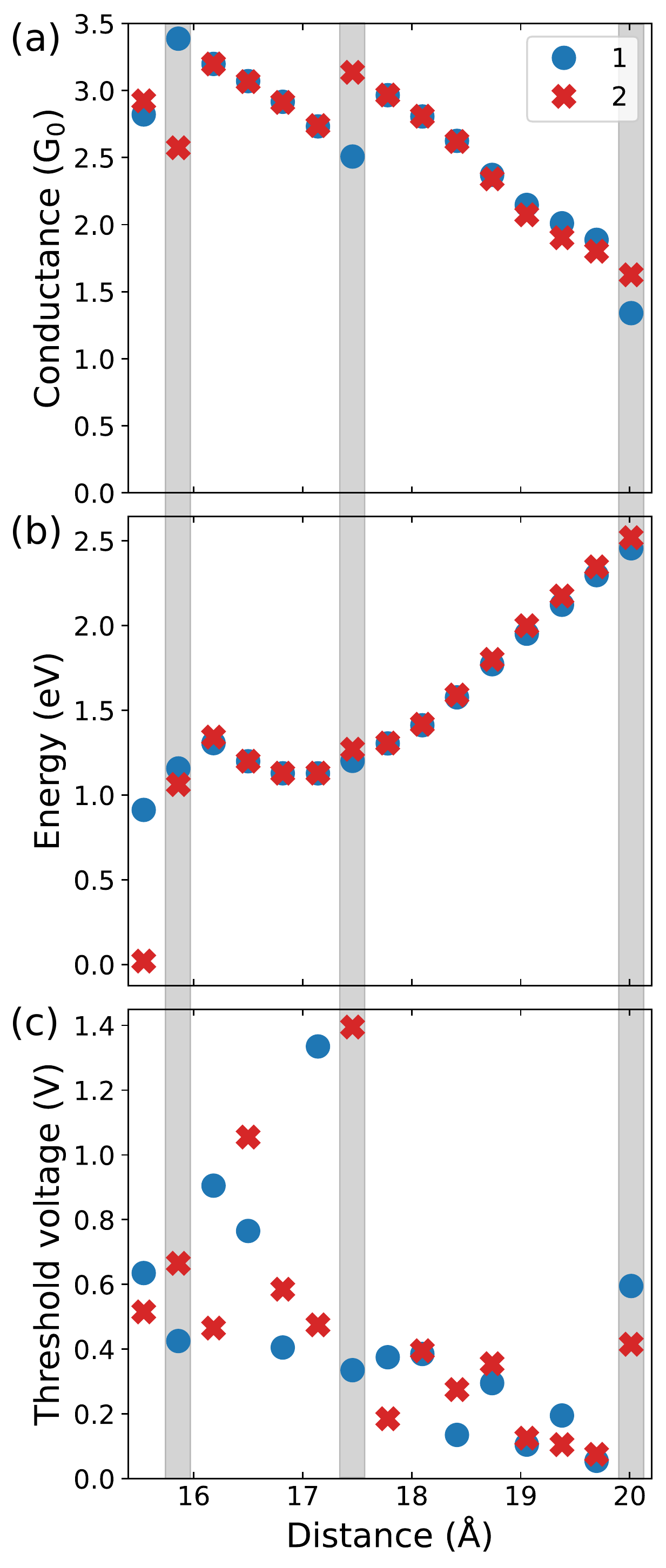}
\caption{Conductance (a), total DFT energy (b) and threshold voltage (c) as a function of electrode separation distance $d$, respectively. Blue points visualize results obtained during an initial compression process. Red points are obtained through a mechanical cycle, by separating the electrodes of the corresponding relaxed junction at the subsequent distance step $d-\Delta$ by one distance step $\Delta=0.3$~\AA\ to reach the electrode separation $d$ and relaxing the contact geometry again. The three gray bars indicate points of bistability in the conductance, as observed in panel (a). 
 \label{fig:3plots}}
\end{figure}

Figure~\ref{fig:3plots} shows the electrical conductances, energies and threshold voltages of a compression process, which started at the largest distance $d=20$~\AA\ (see the Supplemental Material). The conductance increases rather linearly with decreasing distance from around $1.5~G_0$ up to $3.5~G_0$ over a length of about \unit[4]{\AA}. These findings are consistent with earlier calculations \cite{Cuevas1998,Xie2010} and measurements \cite{Scheer1998,Weber2018} for atomic-size Pb wires that exhibit $sp$-orbital conduction with three main transmission eigenchannels at the Fermi energy in a single-atom contact. The conductance curve exhibits three discrete jumps, when considering the initial configurations 1, indicating regions to search for bistable behavior. We generate subsequently the configurations 2 for all electrode separations by the mechanical manipulation cycle of $\mp\Delta$, as explained in Sec.~\ref{sec:Methods}. Electrode separations, where we find bistable conductance behavior, are marked by gray bars in Fig.~\ref{fig:3plots}.
The DFT total energy curve shows a local minimum at around \unit[17]{\AA} and a rather linear slope for higher distance, while the behavior is more complex for shorter distances due to major atomic rearrangements. 
Threshold voltages show a larger spread from 0 to some 1.4~V, with a trend towards decreasing threshold voltages for larger electrode separations. 

\begin{figure}
\includegraphics[width=0.8\columnwidth]{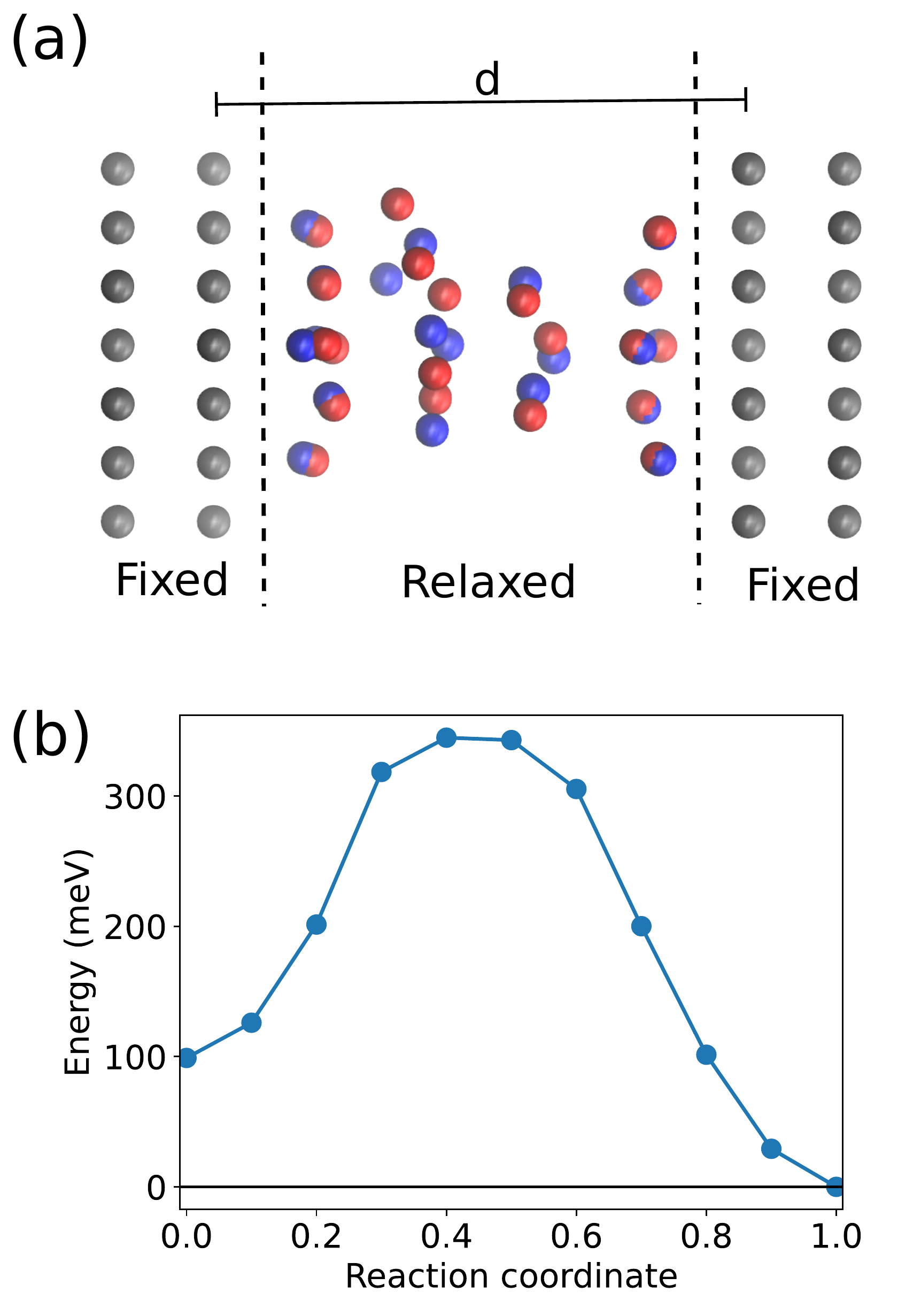}
\caption{(a) Contact configurations 1 (blue) and 2 (red) for the electrode separation $d=$\unit[16]{\AA}. (b) Total DFT energy as a function of the reaction coordinate $r$.  \label{fig:ConfPb14}}
\end{figure}

Figure~\ref{fig:ConfPb14}(a) compares configurations 1 and 2 at $d=16$~\AA. 
We have constructed a linear interpolation between those structures with $10$ steps to analyze the transition. The total DFT energy of the intermediate geometries is shown in Fig.~\ref{fig:ConfPb14}(b), and features a difference in energy of initial and final structures of around \unit[100]{meV}. Depending on a starting point at configuration 1 or 2, the barrier between the structures amounts to around \unit[250-350]{meV}. 

To change these two configurations into each other, it appears that a rotation of a large part of the atoms in the central region is necessary. Unfortunately, we have found no pumped vibrational mode that would enable such a switching. An intuitive explanation for this negative result is that the required rotation does not couple well to the electric current. The charge carriers would need to be scattered nearly orthogonal to their direction of motion, which is unlikely in a two-particle process under momentum conservation. 

\begin{figure}
\includegraphics[width=0.8\columnwidth]{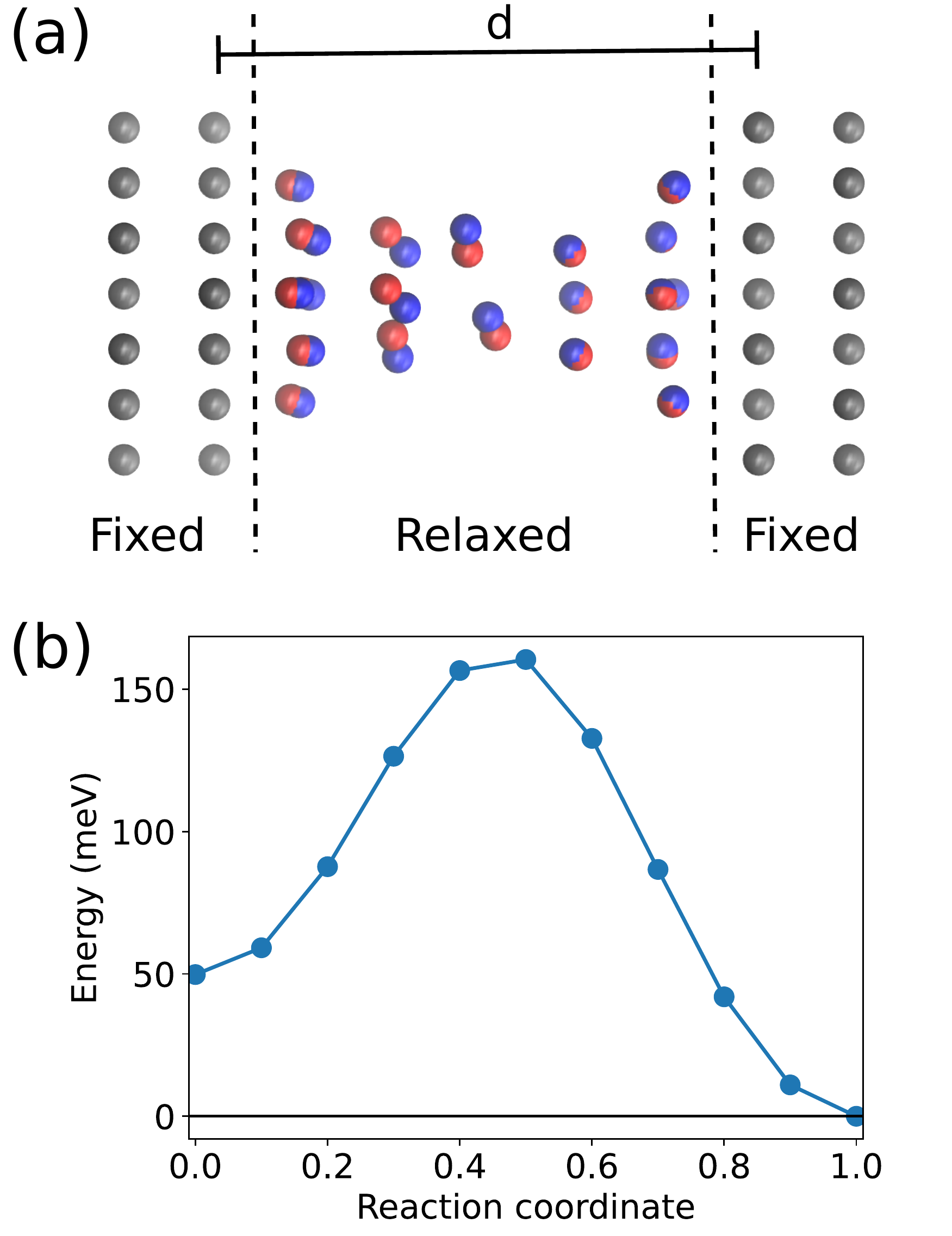}
\caption{Same as Fig.~\ref{fig:ConfPb14} but for $d=$\unit[17.5]{\AA}.  \label{fig:ConfPb11}}
\end{figure}

The switching candidate, shown in Fig.~\ref{fig:ConfPb11}, is also displayed in Fig.~\ref{fig:flowchart}(c)-(e). 
Configurations 1 and 2 are separated by an energy barrier of less than $\unit[160]{meV}$. The blue configuration 1 has a conductance of $2.5~G_0$ and appears to be somewhat more disordered than the red configuration 2, exhibiting a conductance of $3.2~G_0$. Configuration 1 exhibits a higher total energy than configuration 2 by some 50~meV. 
As shown in Fig.~\ref{fig:flowchart}(e), the threshold voltage for pumping vibrational modes amounts to \unit[0.4]{V}, and we can switch to configuration 2 by displacing atoms of configuration 1 along the runaway mode with an amplitude of twice the eigenvector and then relaxing the structure again. In contrast, the threshold voltage in configuration 2 is as high as $\unit[1.4]{V}$, see also \ref{fig:3plots}(c). This indicates a significantly increased stability of that configuration or a significantly reduced electron-vibrational coupling. Unfortunately, we did not find a runaway mode that transforms configuration 2 into configuration 1, and hence this switch is monodirectional.

\begin{figure}
\includegraphics[width=0.8\columnwidth]{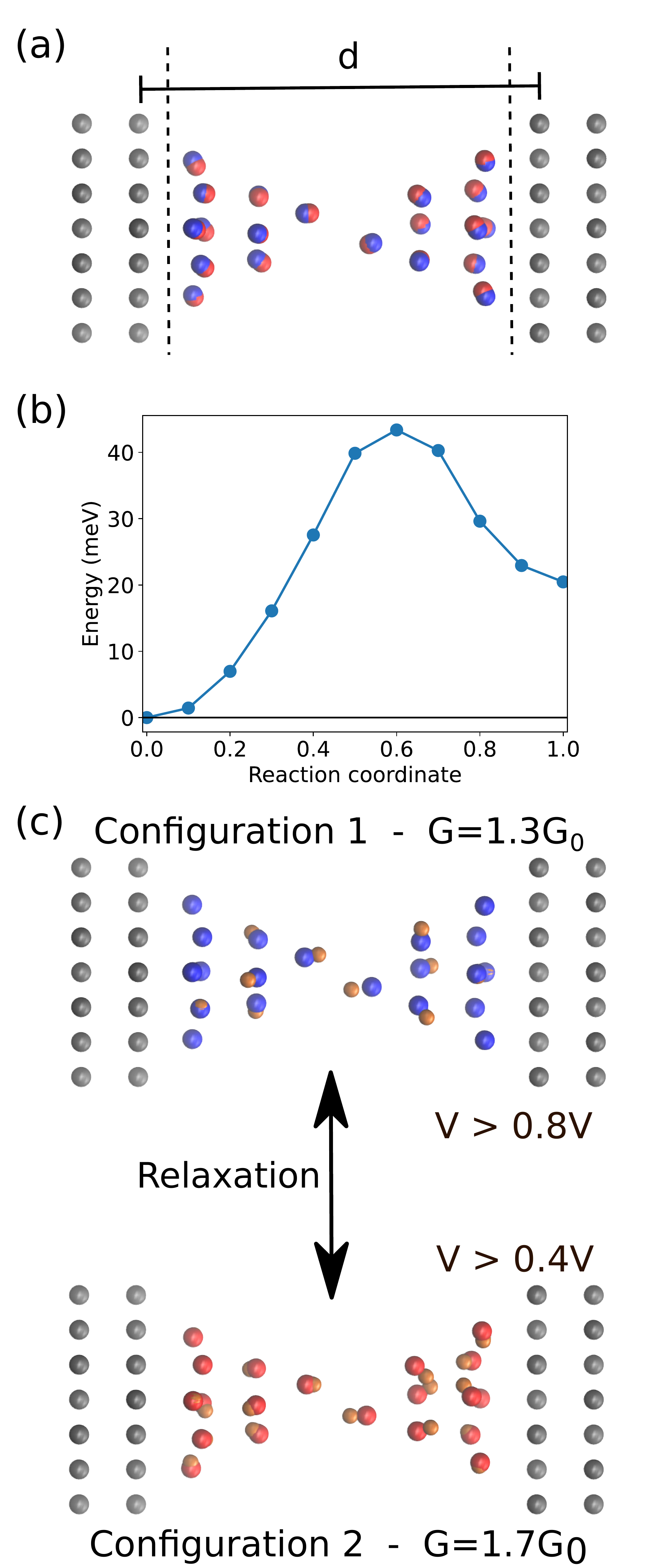}
\caption{(a) Contact configurations 1 (blue) and 2 (red) for $d=$\unit[20]{\AA}. (b) The configurations are separated by a reaction barrier of \unit[20-40]{meV}, depending on the starting point. (c) The contact geometries in panel (a) realize a bistable switch between configuration 1 with a conductance of $1.3~G_0$ and configuration 2 with $1.7~G_0$. Displacements along the pumped vibrational mode are shown in orange, and, according to the stability analysis, require applied bias voltages larger than 0.8 or 0.4~V, respectively. \label{fig:SwitchPb11}}
\end{figure}

Let us finally study the switching candidate at $d=20$~\AA. The atomic configurations of the two states, shown in Fig.~\ref{fig:SwitchPb11}, look very similar. 
The most pronounced relocations are found for the two atoms in the center that form a dimer. The barrier between the configurations has a height of around \unit[40]{meV}, when starting from configuration 1, and \unit[20]{meV}, starting from configuration 2. At sufficiently low temperatures, a crossing of the barrier by thermal excitation alone would be strongly suppressed, while nonconservative current-induced processes should be able to surmount it \cite{Lue2015,Lue2020}. 
For configurations 1 and 2 different runaway vibrational modes could be detected, whose excitation enables the transition into the other configuration. As indicated in Fig.~\ref{fig:SwitchPb11}, the excitation of the mode needs more than \unit[0.8]{V} in configuration 1, while it needs more than \unit[0.4]{V} in configuration 2. Interestingly, the threshold voltages, when transitioning from 1 to 2 instead of 2 to 1, differ by a factor of two, resembling the differences in barrier heights of \unit[40]{meV} and \unit[20]{meV}, respectively. This example of a successful identification of a bistable switch shows the potential of the vibration-mediated switching mechanism. 

Theoretical methods to simulate the switching of atomic contacts by current-induced forces are a timely research theme. The direct molecular dynamics approach offers several challenges, most prevalent the problem of the huge separation in time scales between the current-induced atomic motion and the rare switching events. The approach presented here circumvents that problem by sampling the configurational phase space in the direction of current-induced forces. 
The computational costs of the presented procedure are still significant. The accurate determination of vibrational modes and electron-vibration couplings requires structural relaxations to a high level of convergence. For the switching, many contact configurations are generated and energetically optimized. The high computational demands limit the number of atoms in the extended central cluster of our junction models and of the candidate structures for switches that we could study. For details on computational demands, we refer to the Supplemental Material.

\section*{Conclusions}

In conclusion, we presented a microscopic approach to simulate current-induced switching processes. We used the developed computational method to identify bistable metallic atomic switches, showing two stable atomic configurations with different conductance. The switching is achieved by excitation of a vibrational mode, which becomes amplified by current-induced pumping. Displacing the atoms of the contact along this so-called runaway mode and optimizing atomic positions results in a transition from one contact configuration to the other. Both geometric configurations must be stable over sufficiently long times scales despite the excess energy in the electronic and phononic systems due to the applied bias voltage. This is possible, if the switching process dissipates the excess energy of the respective pumped vibrational modes efficiently, i.e., if the excited mode is not a runaway mode for the new configuration and is thus sufficiently dampened. 

We applied our scheme to study the current-induced reversible switching of Pb nanowires with smallest cross sections containing one or a few atoms only. Combined with our previous findings that computed threshold voltages for a related theoretical approach are of the same size as measured switching voltages  \cite{Ring2020}, our results indicate that the experimentally observed current-induced switching in Al atomic-size contacts \cite{Schirm2013} might also be dominated by electron-vibration scattering. 

In the future, the presented approach could be used to analyze bistable switching in different metals. Furthermore, by repeatedly displacing atoms along runaway modes, it may be possible to study the long-term evolution of contacts towards a higher stability under applied bias. In this way, the microscopic mechanism of electronic hardening may be revealed.

\section*{Acknowledgments}
We thank D. Weber, M. Strohmeier and J. C. Cuevas for inspiring discussions. We gratefully acknowledge financial support by the Deutsche Forschungsgemeinschaft (DFG, German Research Foundation) under project number 262725753,  the Gauss Centre for Supercomputing e.V. (\href{https://www.gauss-centre.eu}{www.gauss-centre.eu})  by providing computing time through the John von Neumann Institute for Computing (NIC) on the GCS Supercomputer JUWELS at the J\"ulich Supercomputing Centre (JSC) \cite{JUWELS}. We acknowledge further computing time provided by the state of Baden-W\"urttemberg through bwHPC and the DFG through project number 236232410 (JUSTUS computing cluster).

\bibliography{paper.bib}

\clearpage


\section*{Supplementary material}

Supplemental Material for
"Simulating bistable current-induced switching of metallic atomic contacts by
electron-vibration scattering"

Markus Ring,[1,2] Fabian Pauly,[1] Peter Nielaba,[2] and Elke Scheer[2]

[1]*Institute of Physics, University of Augsburg, 86159 Augsburg, Germany*
[2]*Department of Physics, University of Konstanz, 78457 Konstanz, Germany*

## S1. MECHANICAL CONTACT MANIPULATION

Figure S1 shows conductances, DFT total energies and threshold voltages for the full simulated compression curve with $d$ ranging from around 10 to 20 Å. In the main text we analyze the contacts at high electrode separations between around 15 and 20 Å, see especially Figs. 1 and 2. We analyze the contacts at large $d$, since the variety of metastable configurations is reduced compared to the smaller distances.

## S2. DEPENDENCE OF RESULTS ON THE EMPLOYED BASIS SET

Figure S2 shows the mechanical manipulation curves, obtained with the high-quality def-TZVP basis set [1, 2]. While conductances and total energies are rather similar to the results for the def-SV(P) basis set, see Figs. 1 and S1, the def-TZVP basis yields threshold voltages closer to experimental values [3]. However, geometry optimizations with the def-TZVP basis set are computationally substantially more demanding. For this reason, we use the def-SV(P) basis set throughout the main text. In general, both basis sets lead to qualitatively similar results, and switches can be discovered independent of the employed basis set via the procedure visualized in Fig. 1.

## S3. COMPUTATIONAL CHALLENGES

In the simulation scheme of Fig. 1 the computationally most time-consuming task is to find the energy-optimized contact configurations of a conductance-distance trace. Only after relaxation at a certain distance $d$ is finished, it is possible to perform the relevant analysis, including the stability analysis of vibrational modes for a flowing electric current. Searching for bistable geometries along the undamped modes requires further relaxations, which take considerably less time. Unfortunately, most of these

searches are unsuccessful, and geometries just return to the initial state, i.e. configuration 1 or 2. Thus, out of several dozen configurations with around five excited modes each, only three modes were found that could lead to a switching into another conductance state.

In summary there is a high initial computational cost of getting the configurations 1 and 2 of each potential switch. The low yield of the runaway-mode-distorted contact geometries to transform configuration 1 into 2 or vice versa is the central problem to extend the proposed scheme to a representative sample size for one or several metals.

## S4. DETAILED COMPUTATIONAL COSTS

The computational costs of the procedure, presented in Fig. 1, are significantly higher than for routine conductance calculations of elastic scattering in a given contact geometry. The vibrational modes and electron-vibration couplings require structure relaxations at a high convergence level. For a specific $d$ the relaxation of the 60-atom Pb contact using the def-SV(P) basis set takes up to several hundred iterations, the typical iteration taking around an hour on a single core on JUWELS [4]. On average a full relaxation at a given electrode separation thus takes $300 - 1000$ core hours, or more than 10.000 core hours for a conductance-distance curve, see Figs. 2 and S1. Computational times increase by a factor of six, when we use the def-TZVP basis set instead of def-SV(P). Therefore, a prerelaxation with def-SV(P) was used to generate Fig. S2. With regard to the switching, we need to generate several candidate structures and relax them after displacing atoms along runaway modes with different amplitudes, see Fig. 1. Those geometries start closer to a known energetic minimum, but the number of relaxation runs (around 20 per candidate switch) again leads to costs of several thousand core hours. The high computational demands constrain the number of atoms in the extended central cluster of the metal junctions and the amount of junction realizations (for instance different stretching or compression curves) that we can acquire with this method.

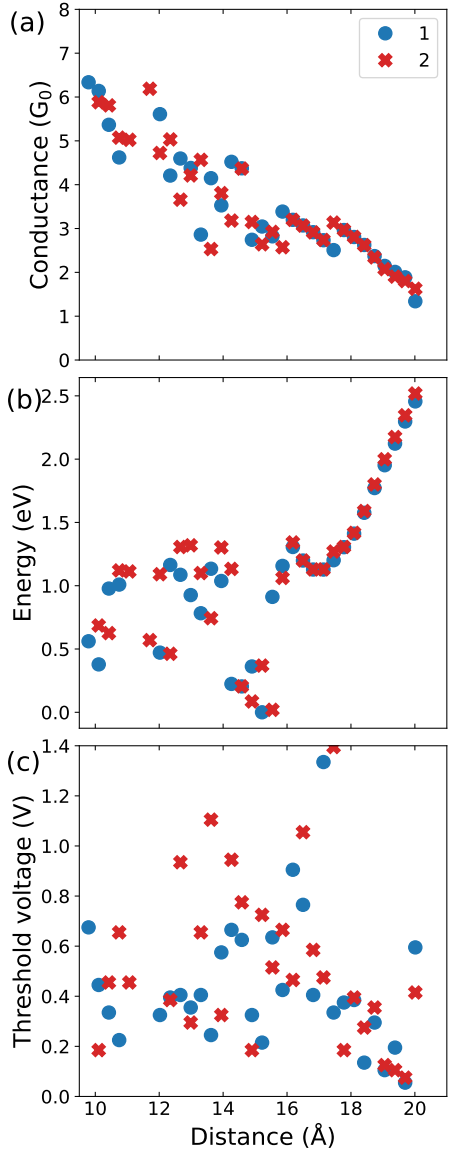

Figure S1. Conductance (a), total DFT energy (b) and threshold voltage (c) as a function of electrode separation distance $d$, respectively. Blue points visualize results obtained during an initial compression process. Red points are obtained through a mechanical cycle, by separating the electrodes of the corresponding relaxed junction at the subsequent distance step $d - \Delta$ by one distance step $\Delta$ to reach the electrode separation $d$ and relaxing the contact geometry again. Figure 2 of the main text shows part of the computed data points at the large distances.

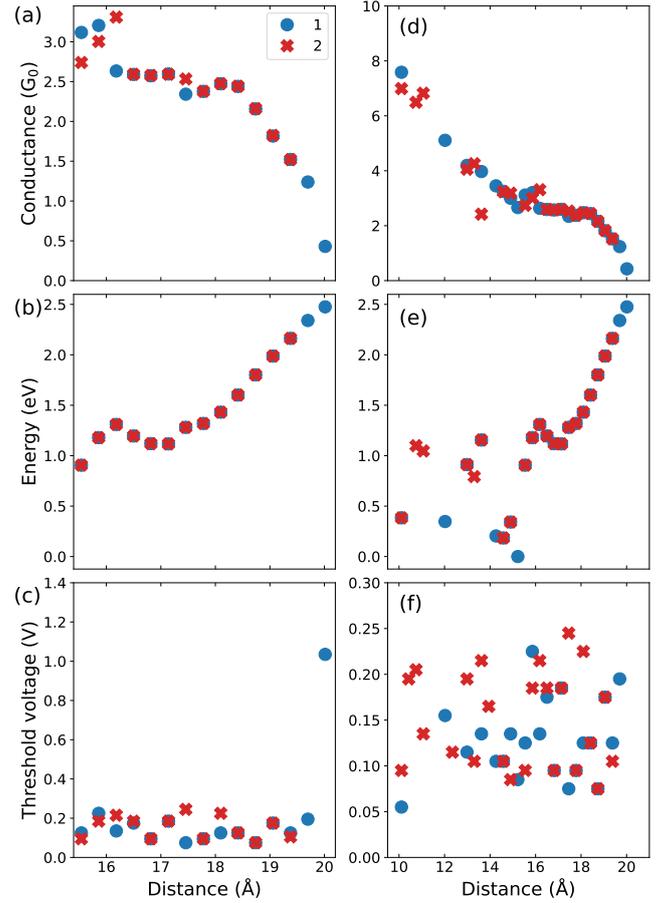

Figure S2. Mechanical contact manipulation, shown in Figs. 1 and S1, determined with the def-TZVP basis set. Conductance (a,d), total DFT energy (b,e) and threshold voltage (c,f) as a function of electrode separation distance $d$, respectively. The left column is a zoom into the full mechanical manipulation, shown in the right column. Blue points visualize results obtained during an initial compression process. Red points are obtained through a mechanical cycle, by separating the electrodes of the corresponding relaxed junction at the subsequent distance step $d - \Delta$ by one distance step $\Delta$ to reach the electrode separation $d$ and relaxing the contact geometry again.



\end{document}